\documentclass[pra]{revtex4}
\usepackage{dcolumn,amsmath}
\usepackage{bm}

\begin{document}

  \renewcommand{\baselinestretch}{1.37}\small\normalsize


%
 \newcommand{\ande}{ {\&} }

 \newcommand{\etale}{ {\sl et~al.}}
 \newcommand{\etal}{ {\sl et~al.}}



 \newcommand{\eref}[1]{(\ref{#1})}
 \newcommand{\eeref}[1]{eq.~(\ref{#1})}
 \newcommand{\erefs}[1]{eqs.~(\ref{#1})}
 \newcommand{\eerefs}[1]{eqs.~(\ref{#1})}
 \newcommand{\frefs}[1]{figures~\ref{#1}}
 \newcommand{\tref}[1]{table~\ref{#1}}
 \newcommand{\trefs}[1]{tables~\ref{#1}}
 \newcommand{\Eref}[1]{Eq.~(\ref{#1})}
 \newcommand{\Erefs}[1]{Eqs.~(\ref{#1})}
 \newcommand{\Frefs}[1]{Figures~\ref{#1}}
 \newcommand{\Tref}[1]{Table~\ref{#1}}
 \newcommand{\Trefs}[1]{Tables~\ref{#1}}
 \newcommand{\Vres}{\ensuremath{V'_{e,e}}}
 \newcommand{\veps}{\ensuremath{\varepsilon}}
 \newcommand{\Aeff}{\ensuremath{A_{\rm eff}}}
 \newcommand{\cm}{\mbox{cm}\ensuremath{^{-1}}} 
 \newcommand{\rtw}{\rightarrow}
 \newcommand{\half}{\ensuremath{\frac{1}{2}}}
 \newcommand{\un}[1]{\underline{\vphantom{pP}#1}}
 \newcommand{\unem}[1]{\underline{\em\vphantom{pP}#1}}
 \newcommand{\Hzecm}{\ensuremath{\frac{\mbox{\normalsize Hz}}
                                      {\mbox{\normalsize e~cm}}}}
 \newcommand{\eHzecm}{\ensuremath{\frac{\mbox{\normalsize Hz}}
                                      {\mbox{\normalsize e~cm}}}}
%
%



 \newcommand{\opH}{{\bf H}}
 \newcommand{\opHEf}{\opH^{\rm Ef}}
 \newcommand{\opHef}{\opH^{\rm Ef}}
 \newcommand{\opHso}{\opH^{\rm SO}}
 \newcommand{\opHSO}{\opH^{\rm SO}}
 \newcommand{\opHo}{\opH^{[0]}} 
 \newcommand{\opHFr}{\opH^{\rm Fr}}
 \newcommand{\opHfr}{\opH^{\rm Fr}}

 \newcommand{\oph}{{\bf h}}
 \newcommand{\opU}{{\bf U}}
 \newcommand{\opUEf}{\opU^{\rm Ef}}
 \newcommand{\opUef}{\opU^{\rm Ef}}
 \newcommand{\opUefHuz}{\opU^{\rm Ef}_{\rm Huz}}
 \newcommand{\opUSfC}{\ensuremath{\ensuremath{\opU}^{\rm SfC}}}
 \newcommand{\opW}{{\bf W}}
 \newcommand{\opF}{{\bf F}}
 \newcommand{\opS}{{\bf S}}

 \newcommand{\oprn}{\vec{r}_{n}}

 \newcommand{\sigp}{(\vec{\sigma}{\cdot}\vec{\bf p})}
 \newcommand{\alphp}{(\vec{\alpha}{\cdot}\vec{\bf p})}
 \newcommand{\opV}{{\bf V}}
 \newcommand{\opVr}{{\it V(r)}}     
 \newcommand{\opVnuc}{{\opV}}
 \newcommand{\opVext}{\opV^{\rm ext}}
 \newcommand{\opVcor}{\opV^{\rm corr}}
 \newcommand{\opVcorr}{\opV^{\rm corr}}

 \newcommand{\opX}{{\bf X}}

 \newcommand{\opJ}{{\bf J}}
 \newcommand{\opK}{{\bf K}}
 \newcommand{\oppsJ}{\widetilde{\bf J}}
 \newcommand{\oppsK}{\widetilde{\bf K}}

 \newcommand{\opP}{{\bf P}}
 \newcommand{\opPl}{\opP_l}
 \newcommand{\opPlj}{\opP_{l\pm}}


 \newcommand{\vecr}{\vec{\it r}}
 \newcommand{\vecp}{\vec{\bf p}}
 \newcommand{\vecalp}{\vec{\alpha}}
 \newcommand{\vecsig}{\vec{\sigma}}


 \newcommand{\OCin}{\mbox{÷ï}\ensuremath{^<}}
 \newcommand{\OMC}{\ensuremath{\mbox{÷ï}^>}}
 \newcommand{\opNomc}{\ensuremath{{\bf N}^>}}
 \newcommand{\Nomc}{\ensuremath{N^>}}
 \newcommand{\DNomc}{\ensuremath{\Delta N}}

 \newcommand{\Rc}{\ensuremath{R_{\rm c}}}
 \newcommand{\Romc}{\ensuremath{r_{\rm omc}}}
 \newcommand{\Rv}{\ensuremath{r_{\rm v}}}
 \newcommand{\Rrest}{\ensuremath{R_C}}

 \newcommand{\Eco}{\ensuremath{E_{\rm core}}}
 \newcommand{\EcoNi}{\ensuremath{\Eco^{N_i}}}
 \newcommand{\dEcoNiDN}{\ensuremath{\delta\EcoNi(\DNomc_i)}}

 \newcommand{\Efr}{\ensuremath{E_{\rm fr}}}
 \newcommand{\dEfr}{\ensuremath{\delta\Efr}}

 \newcommand{\dPsi}{\ensuremath{\delta\Psi}}
 \newcommand{\Psifr}{\ensuremath{\Psi_{\rm fr}}}
 \newcommand{\dPsifr}{\ensuremath{\delta\Psifr}}

 \newcommand{\Esm}{\ensuremath{E_{\rm sm}}}
 \newcommand{\dEsm}{\ensuremath{\delta\Esm}}

 
 \newcommand{\DCBrev}{Grant:02,Visscher:02,Kutzelnigg:02}
 \newcommand{\Kutzeln}{Kutzelnigg:84,Kutzelnigg:89,Kutzelnigg:90,Kutzelnigg:02}
 \newcommand{\KinBal}{Lee:82,Schwarz:82,Schwarz:82b,Dyall:84c}


 \newcommand{\TupTwoAt}{Tupitsyn:00}


 \newcommand{\TwoCompAppr}{Wood:78,Barthelat:80,Lenthe:93,Wolf:02}

 \newcommand{\RCCrev}{Kaldor:99,Landau:01c,Visscher:01}
 \newcommand{\Nether}{Visscher:96bb,DeJong:96,DeJong:99}
 \newcommand{\TAU}{Eliav:96,Eliav:98,Kaldor:99,Landau:01c}
%

 \newcommand{\SODCIp}{Buenker:99,Alekseyev:04,Titov:01}


 \newcommand{\Christ}{Christiansen:79}
 \newcommand{\cErmler}{Ermler:78}
 \newcommand{\cKahn}{Kahn:76}
 \newcommand{\CommentA}{Mosyagin:98cd}
 \newcommand{\MosyaginA}{Mosyagin:94}
 \newcommand{\MosyaginB}{Mosyagin:97}
 \newcommand{\MosyaginHg}{Mosyagin:98bd,Mosyagin:00,Mosyagin:01b,Mosyagin:02}
 \newcommand{\TitovA}{Titov:91}
 \newcommand{\GRECP}{Titov:99,Titov:00,Titov:02Dism,Petrov:03a,Mosyagin:04a}
 \newcommand{\GRThGr}{Titov:99}
 \newcommand{\RECPcomp}{Tupitsyn:95,Mosyagin:97,Titov:99,Titov:00rd,%
Mosyagin:00,Isaev:00,Titov:01,Mosyagin:01b}
 \newcommand{\TupitsynA}{Tupitsyn:95}
 \newcommand{\HFD}{Bratzev:77rd}
 \newcommand{\TupitsynB}{Bratzev:77rd}
 \newcommand{\cBuenker}{Buenker:74,Buenker:75,Buenker:78,Alekseyev:94}

 \newcommand{\PPsolids}{Hamann:79,Bachelet:82,Bachelet:82b,Cohen:83}

 \newcommand{\EAPP}{Kuchle:91,Dolg:92,Haussermann:93,Dolg:93}

 \newcommand{\SepPP}{Blochl:90,Vanderbilt:90,Kresse:99,Pickard:00,Theurich:01}

 \newcommand{\PPsolidsRev}{Heine:73rd,Bachelet:82,Cohen:83,Abarenkov:96rd,%
Hartwigsen:98}

 \newcommand{\CPP}{Fuentealba:83b,Mueller:84a,Mueller:84b,Stoll:84,%
Kuchle:91,Foucrault:92,Magnier:93,Leininger:97}

 \newcommand{\HuzinagaPP}{Bonifacic:74,Katsuki:88b,Seijo:95}

 \newcommand{\RECPrev}{Ermler:88,Schwerdtfeger:03}
 \newcommand{\ECPhistB}{Phillips:59,Abarenkov:65,Weeks:68,Goddard:68,%
Durand:75,Kahn:76,Lee:77,Christiansen:79,Hamann:79,Hafner:79}
 \newcommand{\OCR}{Phillips:59,Pacios:85,Titov:85Dis,Titov:92A,Titov:96,%
Blochl:94}
 \newcommand{\PToddMolCalc}{Kozlov:87,Dmitriev:92,Titov:96b,Kozlov:97,%
Mosyagin:98,Petrov:02,Isaev:04,Petrov:04}

 \newcommand{\PNCrev}{Commins:99,Sapirstein:02aa,Berger:04,Ginges:04}

 \newcommand{\BSSE}{Gutowski:86,Liu:89}
 \newcommand{\GCbasis}{Mosyagin:98bd,Mosyagin:00,Isaev:00,Mosyagin:01b}


 \newcommand{\Tlcalc}{Rakowitz:96,Wahlgren:97,Leininger:97}








 \newcommand{\cHFJ}{HFJ,Tupitsyn:95}
 \newcommand{\cHFDB}{HFDB,Bratzev:77,Tupitsyn:02A}

 \newcommand{\pMOLGEP}{{\sc molgep}}
 \newcommand{\pHFD}{{\sc hfd}}
 \newcommand{\pHFDB}{{\sc hfdb}}
 \newcommand{\pHFDCI}{{\sc hfd/ci}}
 \newcommand{\pHFJ}{{\sc grecp/hfj}}
 \newcommand{\pMOLCAS}{{\sc molcas}}
 \newcommand{\pCCSDT}{{\sc cc-sdt}}
 \newcommand{\pRCC}{{\sc rcc}}
 \newcommand{\pRCCSD}{{\sc rcc-sd}}
 \newcommand{\pRelCCSDT}{{\sc relccsd(t)}}
 \newcommand{\pMRDCI}{{\sc mrd-ci}}
 \newcommand{\pANOL}{{\sc ano-l}}
 \newcommand{\pPTdCI}{{\sc pt2/ci}}

\title{Two-step method for precise calculation of core properties in molecules}

\author{A.V.\ Titov}\email{titov@pnpi.spb.ru}\homepage{http://www.qchem.pnpi.spb.ru}
\author{N.S.\ Mosyagin}
\author{A.N.\ Petrov}
\author{T.A.\ Isaev}

\affiliation{Petersburg Nuclear Physics Institute, 
             Gatchina, St.-Petersburg district 188300, Russia}

\begin{abstract}
 Precise calculations of core properties in heavy-atom systems which are
 described by the operators heavily concentrated in atomic cores, like to
 hyperfine structure and P,T-parity nonconservation effects, usually require
 accounting for relativistic effects.  Unfortunately, completely relativistic
 treatment of molecules containing heavy elements
 is very consuming already at the stages of calculation and transformation of
 two-electron integrals with a basis set of four-component spinors.

 In turn, the relativistic effective core potential (RECP) calculations of
 valence (spectroscopic, chemical etc.) properties of molecules are very
 popular because the RECP method allows one to treat quite satisfactory the
 correlation and relativistic effects for the valence electrons of a molecule
 and to reduce significantly the computational efforts.
 The valence molecular spinors are usually smoothed in atomic cores and, as a
 result, direct calculation of electronic densities near heavy nuclei is
 impossible.

 In the paper, the methods of nonvariational and variational one-center
 restoration of correct shapes of four-component spinors in atomic cores after
 a two-component RECP calculation of a molecule
 are discussed.  Their efficiency is illustrated in correlation calculations of
 hyperfine structure and parity nonconservation effects in heavy-atom molecules
 YbF, BaF, TlF, and PbO.
\end{abstract}

\maketitle

{\bf SHORT NAME:} Studying core properties in molecules

{\bf KEYWORDS FOR INDEXING:} electronic structure, atom in a molecule,
molecules with heavy atoms, method of ab initio molecular calculation,
relativistic effective core potential, one-center restoration.


 \section*{INTRODUCTION}
 \label{sIntro}

\paragraph*{Historical background of study P- and T-parity nonconservation
           effects in heavy-atom molecules.}
 \label{sPThist}

 After discovery of the combined charge (C) and space (P) parity violation, or
 CP-violation, in $K^0$-meson decay \cite{Christenson:64}, the search for the
 electric dipole moments (EDMs) of elementary particles has become one of the
 most fundamental problems in physics
 \cite{Commins:99,Sapirstein:02aa,Berger:04,Ginges:04}.  Permanent EDM is
 induced by the weak interaction that breaks both the space symmetry inversion
 and time-reversal invariance (T) \cite{Landau:57}.  Considerable experimental
 effort has been invested in measuring atomic EDMs induced by the EDMs of
 proton, neutron and electron and by P,T-odd interactions between them.  The
 best available restriction for the electron EDM, $d_e$, was obtained in the
 atomic Tl experiment \cite{Regan:02}, which established an upper limit of
 $|d_e|<1.6\times10^{-27}\ e\cdot$cm, where $e$ is the charge of the electron.
 The benchmark upper limit on a nucleus EDM is obtained in atomic experiment on
 $^{199}$Hg \cite{Romalis:01}, $|d_{\rm Hg}|<2.1\times 10^{-28}\ e\cdot$cm,
 from which the best restriction on the proton EDM, $|d_p|<5.4\times 10^{-24}\
 e\cdot$cm, was also recently obtained by Dmitriev\ande Sen'kov
 \cite{Dmitriev:03}.

 Since 1967, when Sandars suggested to use polar heavy-atom molecules in the
 experimental search for the proton EDM \cite{Sandars:67}, the molecules are
 considered as the most promising objects for such experiments.  Sandars also
 noticed earlier \cite{Sandars:65} that the P- and P,T-parity nonconservation
 (PNC) effects are strongly enhanced in heavy-atom systems due to relativistic
 and other effects.  For example, in paramagnetic atoms the enhancement factor
 for an electron EDM, $d_{\rm atom}/d_e$, is roughly proportional to $\alpha^2
 Z^3 \alpha_D$, where $\alpha \approx 1/137$ is the fine structure constant,
 $Z$ is the nuclear charge and $\alpha_D$ is the atomic polarisability.  It can
 be of order 100 or greater for highly polarizable heavy atoms ($Z\ge50$).
 Furthermore, the effective inner molecular electric field $E_{\rm mol}$ acting
 on electrons in polar molecules can be a few orders of magnitude higher than
 the maximal field $E_{\rm ext}$ accessible in a laboratory, $E_{\rm
 mol}/E_{\rm ext}\sim10^{5}$.
 The first molecular EDM experiment (on search for the proton EDM and other
 nuclear P,T-odd effects) was performed on TlF by Sandars\etal\ \cite{Hinds:76}
 (Oxford, UK).  In 1991, in the last series of the $^{205}$TlF experiments by
 Hinds\etal\ \cite{Cho:91} (Yale, USA), the restriction $d_p =
 (-4\pm6){\times}10^{-23}\ e{\cdot}{\rm cm}$ was obtained (that was
 recalculated in 2002 by Petrov\etal\ \cite{Petrov:02} as $d_p = (-1.7 \pm
 2.8){\times}10^{-23}\ e{\cdot}{\rm cm}$).

 In 1978 the experimental investigation of the electron EDM and other PNC
 effects was further stimulated by Labzowsky\etal\
 \cite{Labzowsky:78,Gorshkov:79} and Sushkov\ande Flambaum \cite{Sushkov:78}
 who clarified the possibilities of additional enhancement of these effects in
 diatomic radicals like BiS and PbF due to the closeness of levels of opposite
 parity in $\Omega$-doublets having the $^2\Pi_{1/2}$ ground state.  Then
 Sushkov\etal\ \cite{Sushkov:84} and Flambaum\ande Khriplovich
 \cite{Flambaum:85b} have suggested to use $\Omega$-doubling in diatomic
 radicals with the $^2\Sigma_{1/2}$ ground state for such experiments and the
 HgF, HgH and BaF molecules were first studied semiempirically by Kozlov
 \cite{Kozlov:85}.  At the same time, the first two-step {\sl ab~initio}
 calculation of PNC effects in PbF initiated by Labzowsky was 
 finished in St.-Petersburg (Russia) \cite{Titov:85Dis}.
 A few years later, Hinds started experimental search for the electron EDM on
 the YbF molecule, on which the first result was obtained by his group in 2002
 (Sussex, UK) \cite{Hudson:02}, $d_e{=}(-0.2 \pm 3.2){\times}10^{-26}
 e{\cdot}{\rm cm}$.  Though that restriction is worse than the best available
 now $d_e$ datum calculated from the Tl experiment (see above), nevertheless,
 it is limited by only counting statistics, as Hinds\etal\ pointed out in
 \cite{Hudson:02}.

 New series of the electron EDM experiments on YbF by Hinds group (London, UK)
 are in progress and new generation of the electron EDM experiments using a
 vapor cell, on the metastable $a(1)$ state of PbO, is prepared by group of
 DeMille (Yale, USA).  The unique suitability of PbO for searching the elusive
 $d_e$ is demonstrated by the very high statistical sensitivity of the Yale
 experiment to the electron EDM.  In prospect, it allows one to detect $d_e$ of
 order of $10^{-31}\ e\cdot$cm \cite{DeMille:00}, three--four orders of
 magnitude lower than the current limit quoted above.
 Some other candidates for the EDM experiments, HgH, HgF, and TeO$^*$, are yet
 discussed and the experiment on PbF is planned in Oklahoma, USA.

 In order to interpret the data measured in such molecular experiments
 in terms of the electron EDM or other fundamental constants of P- and P,T-odd
 interactions, high-precision calculation of the electronic wave function near
 a heavy nucleus is required.
 Moreover, {\sl ab~initio} calculations of some molecular properties are
 usually required already prior to the stage of preparation of the experimental
 setup.

\paragraph*{Heavy-atom molecules, computational strategies.}
 \label{sHAMcalc}

 The most straightforward way for calculation of electronic structure
 of molecules containing heavy atoms is solution of the eigenvalue
 problem using the Dirac-Coulomb or Dirac-Coulomb-Breit Hamiltonian
 \cite{\DCBrev} when some approximation for the four-component wave
 function is chosen.

 However, even applying the relativistic SCF approximation to heavy-atom
 molecules when solving the Dirac-Fock or Dirac-Fock-Breit equations followed
 by the transformation of the two-electron integrals is not always an easy task
 because much larger sets of primitive basis functions are required
 \cite{Visscher:96,Visscher:02aa} for such all-electron four-component
 calculations as compared to the nonrelativistic case.  Starting from the Pauli
 approximation and Foldy\ande Wouthuysen transformation, many different
 two-component approaches were developed in which only large components are
 treated explicitly (e.g., see \cite{\TwoCompAppr} and references).

 The most computational savings are yet achieved when the two-component
 relativistic effective core potential (RECP) approximation proposed by
 Lee\etal\ \cite{Lee:77} is used \cite{\RECPrev}.  There are the following
 reasons for it. The RECP approach allows one to exclude large number of
 chemically inactive electrons from molecular calculations and to treat
 explicitly the valence electrons only.
 Then, the oscillations of the valence spinors are usually smoothed in
 heavy-atom cores.  As a result, the number of primitive basis functions can be
 reduced dramatically.  This approach is based on a well-developed
 nonrelativistic technique of calculations, however, an effective spin-orbit
 interaction and other scalar-relativistic effects are taken into account
 usually by means of radially-local or separable
 potentials~\cite{Ermler:88,Theurich:01}.  Many complications of the
 Dirac-Coulomb(-Breit) molecular calculations \cite{Kutzelnigg:84} are avoided
 when employing RECPs.  The radially-local RECP approaches for
 ``shape-consistent'' (or ``norm-conserving'') pseudoorbitals are the most
 widely applied in calculations of molecules with heavy elements
 though ``energy-adjusted/consistent'' pseudopotentials \cite{Schwerdtfeger:03}
 by Stuttgart team and Huzinaga-type ``{\sl ab~initio} model potentials''
 \cite{Seijo:95} are also actively used.  (In plane~wave calculations of
 many-atomic systems and in molecular dynamics, the separable pseudopotentials
 \cite{\SepPP} are more popular now because they provide linear scaling of
 computational effort on the basis set size.)  The nonrelativistic
 shape-consistent effective core potential has been first proposed by
 Durand\ande Barthelat \cite{Durand:75} and then a modified scheme of the
 pseudoorbital constructing was suggested by Christiansen {\it et al.}
 \cite{Christiansen:79} and by Hamann\etal\ \cite{Hamann:79}.

 However, inaccuracy of the conventional radially-local RECP approaches reaches
 1000--3000\ \cm\ for the transition and dissociation energies that may be
 insufficient in practice.  In deep reorganization of electronic configuration
 structure of molecules containing, in particular, transition metals,
 lanthanides, actinides, and superheavy elements the conventional
 radially-local RECP approaches can not provide reliable results for a wide
 variety of properties (as it was shown both in many calculations and
 theoretically, see \cite{Titov:99,Titov:02Dism} and references) though
 otherwise is sometime stated (e.g., see~\cite{Dolg:00aa,Titov:00b}).
 Such problems can be efficiently overcomed by applying the generalized RECP
 (GRECP) approach \cite{\GRECP}, that involves both radially-local, separable
 and Huzinaga-type \cite{\HuzinagaPP} potentials as its components and as
 particular cases.  In the GRECP concept, the inner core, outer core and
 valence electrons are first treated employing different approximations for
 each.

 Nevertheless, calculation of such properties as electronic
 densities near nuclei, hyperfine structure constants, P,T-parity
 nonconservation effects, chemical and isotopic shifts etc.\ with the help of
 the two-component pseudospinors smoothed in cores is impossible.  For
 evaluation the matrix elements of the operators singular on nuclei, proper
 shapes of the valence molecular four-component spinors must be restored in
 atomic core regions after the RECP calculation of that molecule performed.

 In 1959, a nonrelativistic procedure of restoration of the orbitals from
 smoothed Phillips--Kleinman pseudoorbitals was proposed \cite{Phillips:59}
 based on the orthogonalization of the latters to the original atomic core
 orbitals.  In 1985, Pacios\ande Christiansen \cite{Pacios:85} suggested a
 modified orthogonalization scheme in the case of shape-consistent
 pseudospinors.  At the same time, a simple procedure of one-center restoration
 employing the idea of generation of equivalent basis sets in four-component
 Dirac-Fock and two-component RECP/SCF calculations was proposed in
 \cite{Titov:85Dis} (i.e.\ NOCR procedure, see below) and first applied to
 evaluation of the P,T-odd spin-rotational Hamiltonian parameters in the PbF
 molecule.
 In 1994, similar procedure was used by Bl\"ochl inside the augmentation
 regions \cite{Blochl:94} in solids to construct the transformation operator
 between pseudoorbitals (``PS'') and original orbitals (``AE'') in his
 projector augmented-wave method.

 All the above restoration schemes can be called by ``nonvariational'' as
 compared to the ``variational'' one-center restoration (VOCR, see below)
 procedure proposed in \cite{Titov:92A,Titov:96}.
 Proper behavior of the molecular orbitals (four-component spinors in the
 relativistic case) in atomic cores of molecules can be restored in the scope
 of a variational procedure if the molecular pseudoorbitals (two-component
 pseudospinors) match correctly the original orbitals (large components of
 bispinors) in the valence region after the molecular RECP calculation.  This
 condition is rather correct when the shape-consistent RECP is involved to the
 molecular calculation with explicitly treated outermost core orbitals and,
 especially, when the GRECP operator is used as is demonstrated in
 \cite{Titov:99}.

 At the restoration stage, a one-center expansion on the spherical harmonics
 with numerical radial parts is most appropriate both for orbitals (spinors)
 and for the description of ``external'' interactions with respect to the core
 regions of a considered molecule.  In the scope of the discussed two-step
 methods of the electronic structure calculation of a molecule, finite nucleus
 models and quantum electrodynamic terms including, in particular,
 two-electron Breit interaction \cite{Petrov:03a} may be used without problems.

 One-center expansion had been applied first to whole molecules by
 Desclaux\ande Pyykk\"o in relativistic and nonrelativistic {H}artree-{F}ock
 calculations for the series {CH$_{4}$ to PbH$_{4}$} \cite{Desclaux:74b} and
 then in the Dirac-Fock calculations of CuH, AgH and AuH \cite{Desclaux:76c}
 and other molecules \cite{Desclaux:02}.  A large bond length contraction due
 to the relativistic effects has been estimated.  However, the accuracy of such
 calculations is limited in practice because the orbitals of the hydrogen atom
 are reexpanded on a heavy nucleus in all the coordinate space.  It is worth to
 note that the RECP and one-center expansion approaches were considered earlier
 as alternatives to each other \cite{Pitzer:79,Pyykko:79}.

 The applicability of the proposed two-step algorithms for calculation of wave
 functions of molecules with heavy atoms is
 a consequence of the fact that the valence and core electrons may be
 considered as two subsystems,
 interaction between which is described mainly by some integrated
 and not by detailed properties of these subsystems.
 The methods for consequent calculation of the valence and core parts of
 electronic structure of molecules give us a way to combine relative simplicity
 and accessibility of both molecular RECP calculations in gaussian basis set
 and relativistic finite-difference one-center calculations inside a sphere
 with the atomic core radius.

 The first two-step calculations of the P,T-odd spin-rotational Hamiltonian
 parameters were performed for the PbF radical about 20 years ago
 \cite{Titov:85Dis,Kozlov:87} with the semiempirical accounting for the
 spin-orbit interaction.  Before, only nonrelativistic SCF calculation of the
 TlF molecule using the relativistic scaling was carried out
 \cite{Hinds:80a,Coveney:83}, in which the P,T-odd values were almost three
 times
 underestimated as compared to the relativistic DF calculations. The latter
 were first performed only in 1997 by Laerdahl\etal\ \cite{Laerdahl:97} and by
 Parpia \cite{Parpia:97}. The next two-step calculation, for PbF and HgF
 molecules \cite{Dmitriev:92}, was carried out with
 the spin-orbit RECP part taken into account using the method suggested in
 \cite{Titov:92}.

 Later we performed correlation GRECP/NOCR calculations of the core properties
 in YbF \cite{Titov:96b}, BaF \cite{Kozlov:97}, again in YbF \cite{Mosyagin:98}
 and in TlF \cite{Petrov:02}.  In 1998, first all-electron Dirac-Fock
 calculations of the YbF molecule were also performed by Quiney\etal\
 \cite{Quiney:98} and by Parpia \cite{Parpia:98}.  Very recently we finished
 extensive two-step calculations of the P,T-odd properties and hyperfine
 structure of the excited states of the PbO molecule \cite{Isaev:04,Petrov:04}.

 In the paper, the main features of the used two-step method are presented and
 only the last series of the two-step calculations are discussed, in which
 electron correlations are taken into account by a combined method of the
 second-order perturbation theory (PT2) and configuration interaction (CI), or
 ``PT2/CI'' \cite{Dzuba:96} (for BaF and YbF), by the relativistic coupled
 cluster (RCC) method \cite{Kaldor:97,Landau:01c} (for TlF and PbO), and by the
 spin-orbit direct-CI method \cite{\SODCIp} (for PbO).  In the discussed {\sl
 ab initio} calculations the best up to-date accuracy was attained for the
 hyperfine constants and P,T-odd parameters regarding the molecules containing
 heavy atoms.



\section{Generalized Relativistic Effective Core Potential method.}
 \label{sGRECP}

 When core electrons of a heavy-atom molecule do not play an active role, the
 effective Hamiltonian with RECP can be presented in the form
\begin{equation}
   {\bf H}^{\rm Ef}\ =\ \sum_{i_v} [{\bf h}^{\rm Schr}(i_v) +
          {\bf U}^{\rm Ef}(i_v)] + \sum_{i_v > j_v} \frac{1}{r_{i_v j_v}}\ .
 \label{Heff2}
\end{equation}
 Hamiltonian \eref{Heff2} is written only for a valence subspace of electrons
 which are treated explicitly and denoted by indices $i_v$ and $j_v$.  In
 practice, this subspace is often extended by inclusion of some outermost core
 shells for better accuracy but we will consider them as the valence shells
 below if these outermost core and valence shells are not treated
 using different approximations.
 ${\bf U}^{\rm Ef}$ is an RECP (or relativistic pseudopotential) operator that
 is usually written in the separable (e.g., see \cite{Theurich:01} and
 references) or radially-local (semi-local) \cite{Ermler:88} approximations
 when the valence pseudospinors are smoothed in the heavy-atom cores.  The RECP
 operator simulates, in particular, interactions of the explicitly treated
 electrons with those which are excluded from the RECP calculations.  Besides,
 the generalized RECP operator \cite{Titov:99,Titov:00} described below can be
 used that includes the radially-local, separable and Huzinaga-type
 relativistic pseudopotentials as special cases.  Additionally, the GRECP
 operator can include terms of other types which are important first of all for
 economical but precise treatment of transition metals, lanthanides and
 actinides.  With these terms, accuracy provided by GRECPs can be even higher
 than the accuracy of the frozen core approximation employing the same
 separation on subspaces of valence and core electrons
 \cite{Titov:99,Titov:02Dism}.  In~\Eref{Heff2}, ${\bf h}^{\rm Schr}$ is the
 one-electron Schr\"odinger Hamiltonian
\begin{equation}
     {\bf h}^{\rm Schr}\ = - \frac{1}{2} {\vec \nabla}^2 
     - \frac{Z_{ic}}{r}\ ,
 \label{Schr}
\end{equation}
 where {$Z_{ic}$} is the charge of the nucleus decreased by the number of inner
 core electrons.  Contrary to the four-component wave function used in
 Dirac-Coulomb(-Breit) calculations, the pseudo-wave function in the (G)RECP
 case can be both two- and one-component.  The use of the effective Hamiltonian
 \eref{Heff2} instead of the all-electron relativistic
 Hamiltonians leads to the question about its accuracy.  It was shown both
 theoretically and in many calculations (see \cite{Titov:99} and references)
 that the usual accuracy of the radially-local RECP versions is within
 1000--3000~\cm\ for transition energies between low-lying states.

 The GRECP concept is introduced and developed in a series of papers
 (see \cite{Titov:99,Titov:00,Mosyagin:04a} and references).
 In contrast to other RECP methods, GRECP employs the idea of separating the
 space around a heavy atom into three regions: inner core, outer core and
 valence, which are treated differently.  It allows one to attain practically
 any desired accuracy, while requiring moderate computational efforts.

 The main steps of the scheme of generating the GRECP version with
 the separable correction taken into account are:
\begin{enumerate}
\item   The numerical all-electron relativistic calculation of a generator
        state is carried out for an atom under consideration. For this purpose,
        we use the atomic {\sc hfdb} code~\cite{Bratzev:77,Tupitsyn:02A}.
\item   The numerical pseudospinors  {$\tilde f_{nlj}(r)$} are
        constructed of the large components  {$f_{nlj}(r)$} of the
        {outer core (OC)} and valence (V) HFD(B) spinors so that the innermost
        pseudospinors of them (for each $l$ and $j$) are nodeless, the next
        pseudospinors have one node, and so forth. These pseudospinors satisfy
        the following conditions:
\begin{equation}
 \tilde f_{nlj}(r) =
 \left\{
  \begin{array}{ll}
   f_{nlj}(r),
                                 &  r\geq R_{c}, \\
   y(r)=
   r^{\gamma}\sum_{i=0}^{5}a_{i}r^{i}, &  r<R_{c},
  \end{array}
 \right.
\end{equation}
\[
 \begin{array}{ccc}
  & l=0,1,\ldots,L,~~~~~ j=|l\pm\frac{1}{2}|, & \\
  & n=n_{oc},n_{oc}',\ldots,n_v, &
 \end{array}
\]
        where $n_v, n_{oc}, n_{oc}'$ are principal quantum numbers of the
        valence and outer core spinors,  {$L$} is one more than the highest
        orbital angular momentum of the inner core (IC) spinors. The leading
        power  {$\gamma$} in the polynomial is typically chosen to be close to
        {$L{+}1$} in order to ensure sufficient ejection of the valence and
        outer core electrons from the inner core region.  The $a_i$
        coefficients are determined by the following requirements: 
        \begin{itemize}
	\item    {$\{\tilde f_{nlj}\}$} set is orthonormalized,
        \item    {$y$} and its first four derivatives match
		 {$f_{nlj}$} and its derivatives,
        \item    {$y$} is a smooth and nodeless function, and
        \item    {$\tilde f_{nlj}$} ensures a sufficiently
                smooth shape of the corresponding potential.
	\end{itemize}
         {$R_{c}$} is chosen near the extremum of the spinor so 
	that the corresponding pseudospinor has the defined above number 
	of nodes. In practice, the  {$R_{c}$} radii for the 
	different spinors should be chosen close to each other  
	to generate smooth potentials.
\item   The  {$U_{nlj}$} potentials are derived for each 
        {$l{=}0,\ldots,L$} and  {$j{=}|l \pm \frac{1}{2}|$}        
        for the valence and outer core pseudospinors so that
	the {$\tilde f_{nlj}$} are solutions of
	the nonrelativistic-type Hartree-Fock equations in the
	{\it jj}-coupling scheme for a ``pseudoatom'' with
        the removed inner core electrons.
\begin{eqnarray}
  U_{nlj}(r)  =  \tilde f_{nlj}^{-1}(r)
                \Biggl[\Biggl( \frac{1}{2} 
		{\bf \frac{d^{2}}{dr^{2}} }
                - \frac{l(l+1)}{2r^{2}}
                + \frac{Z_{ic}}{r} 
		 -   \widetilde{\bf J}(r) 
	          +  \widetilde{\bf K}(r)
       	  +  \varepsilon_{nlj}  \Biggr) \tilde f_{nlj}(r)  
	  + \nonumber\\
                \sum_{n'\neq n} \widetilde{\varepsilon}_{n'nlj}
                \tilde f_{n'lj}(r) \Biggr] 
          ,
 \label{U_nlj}
\end{eqnarray}
 where {$\widetilde{\bf J}$} and {$\widetilde{\bf K}$} are the
 Coulomb and exchange operators calculated with the
 {$\tilde f_{nlj}$} pseudospinors, {$\varepsilon_{nlj}$} are the
 one-electron energies of the corresponding spinors, and
 {$\widetilde{\varepsilon}_{n'nlj}$} are off-diagonal Lagrange multipliers
 (which are, in general, slightly different for the original bispinors and
 pseudospinors).
\item The GRECP operator with the separable correction written in the spinor
       representation~\cite{Titov:99} is as
\begin{eqnarray}
 \label{UGRECP}
  {\bf U}^{\rm Ef}  &=&  U_{n_vLJ}(r)
                 +  \sum_{l=0}^L \sum_{j=|l-1/2|}^{l+1/2}
		   \Bigl\{\bigl[U_{n_vlj}(r) 
		    -  U_{n_vLJ}(r)\bigr]
                   {\bf P}_{lj}   \nonumber\\
                &+&   \sum_{n_{oc}} 
                   \bigl[U_{n_{oc},lj}(r) 
		   -  U_{n_vlj}(r)\bigr] 
                   \widetilde{\bf P}_{n_{oc},lj} 
		   +   \sum_{n_{oc}}  \widetilde{\bf P}_{n_{oc},lj}
                   \bigl[U_{n_{oc},lj}(r) 
		    -  U_{n_vlj}(r)\bigr] \\ 
              &-&   \sum_{n_{oc},n_{oc}'} 
                   \widetilde{\bf P}_{n_{oc},lj}
                   \biggl[\frac{U_{n_{oc},lj}(r)+U_{n_{oc}',lj}(r)}{2} 
		      -   U_{n_vlj}(r)\biggr]
                   \widetilde{\bf P}_{n_{oc}',lj}\Bigr\}, \nonumber
\end{eqnarray}
 where
\[
  {\bf P}_{lj} = \sum_{m=-j}^j
    \bigl| ljm \bigl\rangle \bigr\langle ljm \bigr|,
\ \ \ \ \ \ \ \ \ \
  \widetilde{\bf P}_{n_{oc},lj} = \sum_{m=-j}^j
  \bigl| \widetilde{n_{oc},ljm} \bigl\rangle \bigr\langle \widetilde{n_{oc},ljm} \bigr|,
\]
         {$\bigl| ljm \bigl\rangle \bigr\langle ljm \bigr|$} 
 is the projector on the two-component spin-angular function 
	 {$\chi_{ljm}$}, 
         {$\bigl| \widetilde{n_{oc},ljm} \bigl\rangle
                \bigr\langle \widetilde{n_{oc},ljm} \bigr|$}
 is the projector on the outer core pseudospinors 
         {$\tilde f_{n_{oc},lj}\chi_{ljm}$}, 
         and  {$J=L+1/2$}.
\end{enumerate}
 Two of the major features of the GRECP version with the separable correction
 described here are generating of the effective potential components for the
 pseudospinors which may have nodes, and addition of non-local separable terms
 with projectors on the outer core pseudospinors (the second and third lines
 in \Eref{UGRECP}) to the standard semi-local RECP operator (the first line in
 \Eref{UGRECP}).  Some other GRECP versions are described and discussed in
 papers \cite{Titov:99,Titov:00,Titov:02Dism,Mosyagin:04a} in details.

 The GRECP operator in spinor representation \eref{UGRECP} is mainly used in
 our atomic calculations.  The spin-orbit representation of this operator which
 can be found in~\cite{Titov:99} is more efficient in practice being applied to
 molecular calculations.  Despite the complexity of expression \eref{UGRECP}
 for the GRECP operator, the calculation of its one-electron integrals is not
 notably more expensive than that for the case of the conventional
 radially-local RECP operator.


\section{Nonvariational One-Center Restoration method}
 \label{sNOCR}

 The electronic density evaluated from the two-component pseudo-wave function
 (that, in turn, is obtained in the (G)RECP calculation accounting for
 correlation) {very accurately} reproduces the corresponding all-electron
 four-component density in the {valence and outer core regions}.  In the {inner
 core region}, the pseudospinors are smoothed, so that the electronic density
 with the pseudo-wave function is {not correct}.  When operators describing
 properties of interest are heavily concentrated near the nucleus, their mean
 values are strongly affected by the wave function in the inner region.  The
 four-component molecular spinors must, therefore, be restored in the
 heavy-atom cores. 

 All molecular spinors $\phi _{p}$ can be restored as one-center expansions
 in the
 cores using the nonvariational one-center restoration (NOCR)
 scheme \cite{Titov:85Dis,\PToddMolCalc}
 that consists of the following steps:
\begin{itemize}
\item  Generation of {\it equivalent} basis sets of one-center four-component
       spinors \quad\
$
  \left( \begin{array}{c} f_{nlj}(r)\chi_{ljm} \\
     g_{nlj}(r)\chi_{2j{-}l,jm} \\ \end{array} \right)
$
 \quad  and of smoothed two-component pseudospinors
$
    \tilde f_{nlj}(r)\chi _{ljm}
$
 in finite-difference all-electron Dirac-Fock(-Breit) and GRECP/SCF
 calculations of {the same} configurations of a considered atom and its ions.
 The nucleus is usually modeled as a uniform charge distribution within a
 sphere.
 The all-electron four-component {\sc hfdb} \cite{\cHFDB} and two-component
 {\sc  grecp/hfj} \cite{\cHFJ} codes are employed by us to generate two
 equivalent numerical basis sets used at the restoration.  These sets,
 describing mainly the atomic core region, are generated independently of the
 basis set used for the molecular GRECP calculations.

\item
   The molecular {\it pseudospinorbitals} are then expanded in the basis set of
   one-center two-component atomic {pseudospinors}
 (for $r{\le}R_{\rm nocr}$,
 where $R_{\rm nocr}$ is the radius of restoration that should be
 sufficiently large for calculating core properties with required accuracy),
\begin{equation}
    \tilde {\phi} _{p}({\bf x}) \approx
    \sum_{l=0}^{L_{max}}\sum_{j=|l-1/2|}^{j=|l+1/2|} \sum_{n,m}
    c_{nljm}^{p}\tilde f_{nlj}(r)\chi _{ljm}\ ,
 \label{expansion}
\end{equation}
%
 where ${\bf x}$ denotes spatial and spin variables.  Note that for linear
 molecules only one value of $m$ survives in the sum for each ${\phi} _{p}$.

\item
   Finally, the atomic two-component pseudospinors in the molecular basis
   are replaced by equivalent four-component spinors and the expansion
   coefficients from Eq.~(\ref{expansion}) are preserved:
%
\begin{equation} {\phi} _{p}({\bf x}) \approx
    \sum_{l=0}^{L_{\rm max}}\sum_{j=|l-1/2|}^{j=|l+1/2|} \sum_{n,m}
    c_{nljm}^{p}
     \left(
    \begin{array}{c}
    f_{nlj}(r)\chi _{ljm}\\
    g_{nlj}(r)\chi 
    _{2j-l,jm}
    \end{array}
    \right)\ .
 \label{restoration}
\end{equation}
\end{itemize}

 The molecular four-component spinors constructed this way are orthogonal to
 the inner core spinors of the heavy atom, because the atomic basis functions
 used in Eq.~(\ref{restoration}) are generated with the inner core electrons
 treated as frozen.
%


\section{Variational one-center restoration}
 \label{sVOCR}

 In the variational technique of the one-center restoration (VOCR)
 \cite{Titov:92A,Titov:96}, the proper behavior of the four-component molecular
 spinors in the core regions of heavy atoms can be restored as an expansion on
 the spherical harmonics inside the sphere with a restoration radius, $R_{\rm
 vocr}$, that should not be smaller than the matching radius, $R_c$, used at
 the RECP generation.  The outer parts of spinors are treated as frozen after
 the RECP calculation of a considered molecule.  This method enables one to
 combine the advantages of two well-developed approaches, molecular RECP
 calculation on gaussian basis set and atomic-type one-center calculation on
 numerical basis functions, by the most optimal way.  It is considered
 theoretically in \cite{Titov:96} and some results concerning the efficiency of
 the one-center reexpansion of orbitals on another atom can be found in
 \cite{Titov:02Dism}.

 The VOCR scheme can be used for constructing the core parts of the molecular
 spinors (orbitals) instead of using equivalent basis sets as is in the NOCR
 technique.  Some disadvantage of the NOCR scheme is that molecular
 pseudoorbitals are usually computed in a spin-averaged GRECP/SCF molecular
 calculation (i.e.\ without accounting for the effective spin-orbit
 interaction) and the reexpansion of molecular pseudospinorbitals on one-center
 atomic pseudospinors is yet restricted by accuracy, as it was seen in our
 GRECP/RCC/NOCR calculations \cite{Petrov:02} of the TlF molecule (see below).
 With the restored molecular bispinors, the two-electron integrals on molecular
 bispinors can be easily calculated when following the scheme presented
 in~\Erefs{12.1}--\eref{12.4} of the next section (see \cite{Titov:96} for more
 details).  Thus, the four-component transfomation from the atomic basis that
 is now the most time-consuming stage of four-component calculations of
 heavy-atom molecules can be avoided.

 However, the most interesting direction in development of the two-step method
 is the possibility to ``split'' the correlation structure calculation of a
 molecule on two consequent correlation calculations: first, in the valence
 region, when the outer core and valence electrons are explicitly involved into
 the GRECP calculation, and then, in the core region, when the outer and
 inner core space regions are only treated at the restoration stage.
 Correlation effects in the valence and outer core regions (not only valence
 parts of molecular orbitals but also configuration coefficients) can be
 calculated, for example, by the GRECP/RCC method with very high accuracy.
 Then correlation effects in the inner and outer core regions can be taken into
 account using the Dirac-Coulomb(-Breit) Hamiltonian and the one-center
 expansion.  Increasing the one-center restoration radius $R_{\rm vocr}$ one
 can take into account correlation effects in the intermediate region (outer
 core in our case) with the required accuracy.
 Roughly speaking, the computational efforts for the correlation structure
 calculations in the core and valence regions are added when using the two-step
 approach, whereas in the conventional one-step scheme, they have
 multiplicative nature.



\section{Two-step calculation of molecular properties.}
 \label{s2stProp}

 To evaluate one-electron core properties (hyperfine structure, P,T-odd effects
 etc.) employing the above restoraton schemes it is sufficient to obtain the
 one-particle density matrix, $\{\widetilde{D_{pq}}\}$, after the molecular
 RECP calculation on the basis of pseudospinors $\{\widetilde{\phi}_p\}$.  At
 the same time, the matrix elements $\{{W}_{pq}\}$ of the property operator
 $\bm{W}({\bf{x}})$ should be calculated on the basis of equivalent
 four-component spinors $\{{\phi}_p\}$.  The mean value for this operator can
 be then evaluated as
\begin{equation} 
   \langle {\bm{W}} \rangle\ =\ \sum_{pq} \widetilde{D_{pq}} {W}_{pq}\ .
 \label{<W>} 
\end{equation} 
 The only explicitly treated valence shells are used for evaluating a core
 property when applying \Eref{<W>} since the atomic frozen core (closed)
 shells do not usually contibute to the properties of practical interest.
 However, correlations with the core electrons explicitly excluded from the
 RECP calculation
 can be also taken into account if the effective operator technique
 \cite{Lindgren:84} is applied to calculate the property matrix elements
 $\{{W}_{pq}^{\rm Ef}\}$ on the basis set of bispinors $\{{\phi}_p\}$.  Then,
 in expression \eref{<W>} one should only replace $\{{W}_{pq}\}$ by
 $\{{W}_{pq}^{\rm Ef}\}$.  
 Alternatively, the correlations with the inner core electrons can be, in
 principle, considered within the variational restoration scheme for electronic
 structure in the heavy atom cores.
 Strictly speaking, the use of the effective operators is correct when the
 molecular calculation is carried out with the ``correlated'' GRECP (see
 \cite{Mosyagin:04a}), in which the same correlations with the excluded core
 electrons are taken into account at the GRECP generation as they are in
 constructing $\{{W}_{pq}^{\rm Ef}\}$.  Nevertheless, the use of the (G)RECP
 that does not account for the core correlations at the molecular calculation
 stage can be justified if they do not influence dramatically on the electronic
 structure in the valence region.  The latter approximation was applied in
 calculations of the YbF and BaF molecules described in the following section.

\vspace{3mm}
 When contributions to $\langle {\bm{W}} \rangle$ are important both in the
 core and valence regions, the scheme for evaluating the mean value of
 $\bm{W}(\bf{x})$ can be as follows:
\begin{itemize}
\item 
 calculation of matrix elements on the molecular pseudospinorbitals (which are
 usually linear combination of atomic gaussians) over all the space region
 using conventional codes for molecular RECP calculations (though, the operator
 $\bm{W}$ can be presented in different forms in calculations with the RECP and
 Dirac-Coulomb(-Breit) Hamiltonians),
\begin{equation}
        \widetilde{\langle \bm{W} \rangle}\ =\ 
        \sum_{pq} \widetilde{D_{pq}}
                \int\limits_{r< \infty}
                \widetilde{\phi}_p({\bf{x}})\
                \bm{W}({\bf{x}})\
                \widetilde{\phi}_q({\bf{x}})\ d{\bf{x}}\ ;
 \label{12.1}
\end{equation} 
\item
 calculation of the inner part of the matrix element of the operator with the
 same molecular pseudospinorbitals using the one-center expansion for
 $\{\widetilde{\phi}_p\}$
 ($R_{\rm ocr}$ stays for $R_{\rm nocr}$ or $R_{\rm vocr}$ below,
 $R_{\rm ocr} \ge R_c$):
\begin{equation}
        {\widetilde{\langle \bm{W} \rangle}}^<\ =\ 
        \sum_{pq} \widetilde{D_{pq}}
                \int\limits_{ r < R_{\rm ocr}}
                \widetilde{\phi}_p({\bf{x}})\
                {\bm{W}}({\bf{x}})\
                \widetilde{\phi}_q({\bf{x}})\ d{\bf{x}}\ ;
 \label{12.2}
\end{equation} 
\item
 calculation of the inner part of the matrix element of the operator with the
 restored molecular four-component spinors using the one-center expansion for
 $\{{\phi}_p\}$:
\begin{equation}
        {\langle \bm{W} \rangle}^<\ =\ 
        \sum_{pq} \widetilde{D_{pq}}
                \int\limits_{r < R_{\rm ocr}}
                \phi^<_p (\bf{x})\
                {\bm{W}} (\bf{x})\
                \phi^<_q (\bf{x}) \ d\bf{x}\ .
 \label{12.3}
\end{equation}
\end{itemize} 
 The matrix element $\langle \bm{W} \rangle$ of the 
 $\bm{W}(\bf{x})$ operator is evaluated as
\begin{equation}
        {\langle \bm{W} \rangle}\ =\ \widetilde{\langle \bm{W}
                 \rangle}\ -\
                {\widetilde{\langle \bm{W} \rangle}}^<\
                +\ {\langle \bm{W} \rangle}^<\ .\
 \label{12.4}
\end{equation}
 Obviously, that the one-center basis functions can be numerical
 (finite-difference) for better flexibility.

 The mean values of the majority of operators for the valence properties
 can be calculated with good accuracy without accounting for the inner parts of
 the matrix elements \eref{12.2} and \eref{12.3}.  As is discussed above, for
 calculating the mean values of the operators singular near nuclei
 it is sufficient to account only for the inner parts, \eref{12.3}, of the
 matrix elements of $\bm{W}(\bf{x})$ on the restored functions
 $\phi^<_p({\bf{x}})$, whereas the other contributions, \eref{12.1} and
 \eref{12.2}, can be neglected.

 Calculation of properties using the finite-field method
 \cite{Kunik:71,Monkhorst:77} and \Eref{<W>} within the coupled-cluster
 approach is described in section \ref{sTlF}.








\section{Calculation of spin-rotational Hamiltonian parameters 
         for YbF and BaF molecules}
 \label{sBaFYbF}

\paragraph*{Effective Hamiltonian.}
 \label{sBaFYbFHeff}

 The molecular spin-rotational degrees of freedom of YbF and BaF for
 the $^{171}$Yb and $^{137}$Ba isotopes with nuclear spin $I{=}\frac{1}{2}$
 are described by Hamiltonian \cite{Kozlov:87,Dmitriev:92,Kozlov:95}
\begin{eqnarray}
        {\bf H}_{\rm sr} & = & B\vec{\bf N}^2 + \gamma \vec{\bf S} \vec{\bf N}
                 - D_e\vec{\lambda} \vec{E}
                 + \vec{\bf S} \hat{\bf A} \vec{\bf I}
\nonumber\\
        & + & W_{\rm A} k_{\rm A} \vec{\lambda}{\times}\vec{\bf S} \cdot
                             \vec{\bf I}
                         +(W_{\rm S} k_{\rm S}
                         + W_d d_{\rm e}) \vec{\bf S}\vec{\lambda}\ ,
 \label{eYbF1}
\end{eqnarray}
 where $\vec{\bf N}$ is the rotational angular momentum, $B$ is the
 rotational constant, $\vec{\bf S}$ is the effective spin of the electron
 \cite{Kozlov:95}, $\vec{\bf I}$ is the spin of the $^{171}$Yb ($^{137}$Ba) 
 nucleus, $\vec{\lambda}$ is the unit vector directed along the molecular
 axis from the heavy nucleus to fluorine.
 The spin-doubling constant $\gamma$ characterizes the spin-rotational
 interaction.  $D_e$ and $\vec{E}$ are the molecular dipole moment and
 the external electric field. The axial tensor $\hat{\bf A}$ describes
 magnetic hyperfine structure on the ytterbium (barium) nucleus:
\begin{equation}
  {\bf H}_{\rm hfs} = \frac{\mu_N}{I} \frac{\vec{\bf I}\cdot
                       \vec{\alpha}{\times}\vec{r}}{r^3}\ ,
   \qquad \vec{\alpha} =
     \left(\begin{array}{cc}
        0 & \vec{\sigma}\\
        \vec{\sigma} & 0
     \end{array} \right)\ ,
 \label{eYbF2}
\end{equation}
 which can be also expressed in the notations of the
 isotropic $A{=}(A_{\parallel}{+}2A_{\perp})/3$ and dipole $A_{\rm
 d}{=}(A_{\parallel}{-}A_{\perp})/3$ hyperfine structure constants.
 The hyperfine interaction with the $^{19}$F nucleus is significantly
 smaller and is not of interest for the considered effects.

 The last three terms in \Eref{eYbF1} account for the P- and
 P,T-odd effects:
 {\em the first} of them describes the electromagnetic interaction of the
 electron with the anapole moment of the nucleus with the constant
 $k_{\rm A}$ \cite{Khriplovich:91};
 {\em the second} term accounts for the scalar P,T-odd electron-nucleon
 interaction characterized by the dimensionless constant $k_{\rm S}$;
 {\em the third} term corresponds to interaction of the electron EDM $d_{\rm
 e}$ with the internal molecular field $\vec{E}_{\rm mol}$:
\begin{eqnarray}
        &&{\bf H}_{d} = 2 d_{\rm e}
        \left(\begin{array}{cc}
        0 & 0 \\
        0 & \vec{\sigma}
        \end{array} \right)
        \cdot \vec{E}_{\rm mol}\ ,
 \label{eYbF1a}\\
        &&W_d d_{\rm e} =
        2 \langle ^2\Sigma_{1/2}|\sum_i {\bf H}_{d}(i)| ^2\Sigma_{1/2} \rangle\ .
 \label{eYbF1b}
\end{eqnarray}
 The constant $\frac{1}{2}W_d$ characterizes an effective electric field on the
 unpaired electron. All the P- and P,T-odd constants $W_X$ depend on the
 electron spin density in the vicinity of the heavy nucleus.  The reliability
 of their calculation can be tested by comparison of the calculated and
 experimental values for the hyperfine constants $A$ and $A_d$.

\paragraph*{Details of calculations.}
 \label{sBaFYbFcalc}

 Let us consider the scheme of calculation for the above molecules on example
 of the YbF calculation carried out mainly in paper \cite{Mosyagin:98}.  The
 stage of the GRECP calculation of the valence structure for the ground state
 $^2\Sigma$ of YbF was performed by analogy with the previous YbF calculation
 \cite{Titov:96b}. The main difference from the latter was the use of technique
 \cite{Titov:99,Titov:01} for freezing the $5s$ and $5p$ pseudospinors derived
 from the calculation on the Yb$^{2+}$ cation. It was necessary to freeze these
 shells because their polarization was taken into account at the stage of
 calculating the effective operator (EO, see~\cite{Dzuba:96,Kozlov:97b} for
 details).
 The spin-averaged GRECP calculations in the framework of the RASSCF method
 \cite{Olsen:88,MOLCAS} with 5284 configurations were performed for 11
 electrons distributed in RAS-1$=$(2,0,0,0), RAS-2$=$(2,1,1,0) and
 RAS-3$=$(6,4,4,2) active orbital subspaces according to the $A_1,B_1,B_2$ and
 $A_2$ irreducible representations (irreps) of the $C_{2v}$ symmetry group used
 in the calculation (see paper \cite{Olsen:88} for details on the RASSCF
 method). Similar calculation was also carried out for the BaF
 molecule~\cite{Kozlov:97}.

 Expressions for the $W_d$ parameter which characterizes the internal 
 molecular field are presented in \Erefs{eYbF1a} and \eref{eYbF1b}, 
 whereas expressions for other electronic matrix elements which correspond 
 to the parameters $A$, $A_{\rm d}$ and $W_X$ of operator~\eref{eYbF1} can 
 be found in papers~\cite{Kozlov:87,Kozlov:95}.  
 All the radial integrals and atomic four-component spinors were calculated for
 the finite nucleus $^{171}$Yb using a model of a uniform charge distribution
 within a sphere. As is mentioned in the introduction, atomic matrix elements
 of operator~\eref{eYbF1a} are proportional to $Z^3$. The same scaling is
 applicable to the constant $W_{\rm S}$, while the matrix elements which
 contribute to the constants $W_{\rm A}$ are proportional to $Z^2$. The nuclear
 charge of fluorine is eight times smaller than that of ytterbium, therefore,
 the contributions to the $W_X$ parameters from the vicinity of the fluorine
 nucleus can be neglected. The additional argument in favour of this
 approximation is connected with the circumstance that the state of the
 unpaired electron in the YbF molecule is described mainly by the ytterbium
 orbitals and thus the spin density is also localized near ytterbium.

\paragraph*{Results.}
 \label{sBaFYbFres}

 Results of the calculation for the spin-rotational Hamiltonian parameters are
 presented in \Tref{tBaFYbF}.  Comparison of the results of the
 GRECP/RASSCF calculations \cite{Titov:96b} with the results of the
 GRECP/RASSCF/EO calculations \cite{Mosyagin:98} confirms the conclusion made
 in paper \cite{Titov:96b} that the core-valence correlations of the unpaired
 YbF electron occupying mainly the hybridized $6s-6p$ orbital of Yb with the
 $5s$ and then with $5p$ shells of the Yb core (first of all, spin polarization
 of the $5s,5p$ shells) contribute very essentially to the hyperfine
 and the P- and P,T-odd constants.

 Our final results for the hyperfine structure constant $A$ differ by less than
 3\% from the experimental value~\cite{Knight:70}. This means that the
 space-isotropic part of the unpaired spin density in the vicinity of the Yb
 nucleus in our calculations is obtained with rather good accuracy.  However,
 it is also important to reproduce the anisotropic (dipole) part of the spin
 density which contributes to the dipole constant $A_{\rm d}$.  The value for
 $A_{\rm d}$ obtained in the RASSCF/EO calculation is in slightly better
 agreement with the experimental datum than the value from paper
 \cite{Quiney:98} but is still underestimated by 23\%.  About one half of this
 difference can be explained by the fact that the $4f$ shell of Yb was frozen
 in our molecular calculation (it was first pointed out by Khriplovich that
 accounting for excitations from the $4f$-shell can be important).  In
 particular, these excitations can explain the small value of the spin-doubling
 constant $\gamma$ \cite{Sauer:95,Sauer:96}.  It was shown in semiempirical
 calculation \cite{Kozlov:97b} that the contribution of the $4f$-shell
 excitations to the spin density can result in a significant correction to the
 constant $A_{\rm d}$.  Using Eqs.~(19) and (31) from paper
 \cite{Kozlov:97b}, one can go to the following estimates for the contributions
 from the $4f$-shell excitations to $A$ and $A_{\rm d}$:
\begin{equation}
        \delta A \approx -3 \mbox{ MHz}, \qquad
        \delta A_{\rm d} \approx 15 \mbox{ MHz}.
\label{eYbF3.2}
\end{equation}
 Note that the semiempirical corrections obtained in \cite{Mosyagin:98}
 arise from the admixture to the molecular wavefunction of the
 configuration with the hole in the $4f$-shell. The weight of this
 configuration was estimated in paper \cite{Kozlov:97b} to be of the order
 of 4\%. This admixture is a purely molecular effect and is not described 
 by the effective operator technique.  Thus, one can conclude that the 
 $\delta A_{\rm d}$ value can be added to the RASSCF/EO value for $A_{\rm d}$,
 resulting in $A_{\rm d} \approx 94$~MHz which is essentially closer to
 the experimental value of 102~MHz.
 Note also that {\sl ab initio} correlation calculation of YbF with, at least,
 $4f$ shell treated explicitly should be performed to check the above described
 mechanism of increasing the calculated $A_{\rm d}$ value.  It is likely that
 the real situation is more complicated and indirect contributions to $A_{\rm
 d}$ caused by the $4f$ shell correlation--relaxation effects and by other
 higher order contributions (which are not accounted for in the considered
 calculation) are mainly responsible for the obtained underestimation of
 $A_{\rm d}$.

 It is essential that similar contribution of the $4f$-shell excitations to the
 constant $W_d$ is strongly suppressed.  Indeed, operator \eref{eYbF1a} mixes
 $f$- and $d$-waves.  The $4d$-shell is localized rather deeply in the
 ytterbium core and its mixture with the $4f$-orbitals by the molecular field
 is very small while the $5d$ electrons are relatively weakly-bound and does
 not penetrate essentially into the region of the $4f$-shell localization.
 Similar contributions to other constants $W_X$ are negligible due to the
 contact character of the corresponding interactions.

 One can see that the values of the $W_d$ constant from the unrestricted DHF
 calculation (accounting for the spin-polarization of core shells)
 \cite{Parpia:98}, the latest semiempirical calculation~\cite{Kozlov:97b} and
 the GRECP/RASSCF/NOCR/EO calculation are in very good agreement.  It is also
 important for reliability of the carried out calculations that the
 contribution from the valence electron to $W_d$ in paper \cite{Parpia:98}
 differs only by 7.4\% from the corresponding GRECP/SCF/NOCR calculation
 \cite{Titov:96b} (see \Tref{tBaFYbF}).  Another DHF calculation
 \cite{Quiney:98} gives a value that is about two times smaller
 (that is rather explained by different definitions used for $W_d$).

 Similar increase in the values for the hyperfine constants and parameters of
 the P,T-odd interactions when the correlations with the core shells (first of
 all, $5s,5p$) are taken into account is also observed for the BaF molecule
 \cite{Kozlov:97} as one can see in \Tref{tBaFYbF}).  Of course, the
 corrections on the $4f$-electron excitations are not required for this
 molecule.  The enhancement factor for the P,T-odd effects in the BaF molecule
 is three times smaller than in YbF mainly because of smaller nuclear charge of
 Ba.  Therefore, barium fluoride can be considered as a less promising molecule
 for a search for the PNC effects.

\begin{table*}
\caption{Parameters of the spin-rotational Hamiltonian for
         the $^{171}$YbF and $^{137}$BaF molecules;  
	 $A=(A_{\parallel}+2A_{\perp})/3$ (isotropic) and
         $A_d=(A_{\parallel}-A_{\perp})/3$ (dipole) are the HFS constants.}
 \label{tBaFYbF}

\begin{tabular}{lcclcl}
\hline
\hline
\vspace{-4mm}\\
                &  $A$  &$A_{d}$&        \quad  $W_d$                & $W_A$ & $W_S$\\
                              & (MHz) & (MHz) & ($10^{25}$~\Hzecm) & (kHz) & (Hz)\\
\vspace{-4mm}\\
\hline
\hline
\multicolumn{6}{c}{\bf The $\phantom{\Bigl(}^{171}$YbF molecule} \\
\hline
\vspace{-3mm}\\
 Experiment~\cite{Knight:70}
                                        & 7617 & 102 &          &     &      \\
 Semiempirical~\cite{Kozlov:94}             &      &     & $-$1.5   & 730 & $-$48\\
 Semiempirical~\cite{Kozlov:97}
                     (with $4f$-correction) &      &     & $-$1.26  &     & $-$43\\
\hline
 GRECP/SCF/NOCR \cite{Titov:96b}        & 4932 & ~59 & $-$0.91  & 484 & $-$33\\
 GRECP/RASSCF/NOCR \cite{Titov:96b}     & 4854 & ~60 & $-$0.91  & 486 & $-$33\\
\hline
 Restricted DHF (Quiney, 1998) \cite{Quiney:98}
                                        & 5918 & ~35 & $-$0.31  & 163 & $-$11\\
 Restricted DHF + core polarization         & 7865 & ~60 & $-$0.60  & 310 & $-$21\\
 Rescaled restricted DHF                    &      &     & $-$0.62  & 326 & $-$22\\
\hline
 Unrestricted DHF (Parpia, 1998) \cite{Parpia:98}\\
 ~~~ (unpaired valence electron)        &      &     & $-$0.962 &     &      \\
 Unrestricted DHF \cite{Parpia:98}      &      &     & $-$1.203 &     & $-$22\\
\hline
 GRECP/RASSCF/NOCR/EO \cite{Mosyagin:98}& 7842 & ~79 & $-$1.206 & 634 &      \\
 GRECP/RASSCF/NOCR/EO \cite{Mosyagin:98}\\
 ~~~         (with $4f$-correction) & 7839 & ~94 & $-$1.206 & 634 &  \\
\hline
\hline
\vspace{-4mm}\\
\multicolumn{6}{c}{\bf The $\phantom{\Bigl(}^{137}$BaF molecule} \\
\hline
\vspace{-3mm}\\
 Experiment~\cite{Knight:71}\footnotemark[1]
                                        & 2326 & 25 &          &     &       \\
 Semiempirical~\cite{Kozlov:85}
                                        &      &    & $-$0.41  & 240 &$-$13  \\
 Experiment~\cite{Ryzlewicz:82}\footnotemark[2]
                                        & 2418 & 17 &          &     &       \\
 Semiempirical~\cite{Kozlov:85}
                                        &      &    & $-$0.35  & 210 &$-$11  \\
\vspace{-4mm}\\
\hline
\vspace{-4mm}\\
 GRECP/SCF/NOCR \cite{Kozlov:97}        & 1457 & 11 & $-$0.230 & 111 &~$-$6.1\\
 GRECP/RASSCF/NOCR \cite{Kozlov:97}     & 1466 & 11 & $-$0.224 & 107 &~$-$5.9\\
 GRECP/SCF/NOCR/EO \cite{Kozlov:97}     & 2212 & 26 & $-$0.375 & 181 &       \\
 GRECP/RASSCF/NOCR/EO \cite{Kozlov:97}  & 2224 & 24 & $-$0.364 & 175 &       \\
\hline
\hline
\end{tabular}

\hspace{-3mm}
\footnotetext[1]{The HFS constants are measured
                in an inert gas matrix \cite{Knight:71} and
                the semiempirical
                $W_X$ values \cite{Kozlov:85} are based on these data.}
\hspace{-3mm}
\footnotetext[2]{The HFS constants were measured in a molecular beam 
                \cite{Ryzlewicz:82}.}

\end{table*}


\section{Calculation of~ $^{205}$TlF molecule.}
\label{sTlF}

\paragraph*{Effective Hamiltonian.}
 \label{sTlFHeff}

 Here we consider the interaction of the EDM of unpaired proton in $^{205}$Tl
 with the internal electromagnetic field of the $^{205}$TlF molecule
 \cite{Petrov:02}.  This molecule is one of the best objects for the proton
 EDM measurements.  The effective interaction with the proton EDM in TlF is
 written in the form
\begin{equation}
     H^{\rm eff}=(d^V+d^M) \vec{I}/I \cdot \vec{\lambda}\ ,
 \label{interaction}
\end{equation}
 where $\vec{I}$ is the Tl nuclear spin operator,
 $\vec{\lambda}$ is the unit vector along $z$ (from Tl to F),
 $d^V$ and $d^M$ are {\it volume} and {\it magnetic} constants
 \cite{Schiff:63}
\begin{equation}
   d^V=6SX=(-d_pR+Q)X\ ,
\vspace{3mm}
 \label{dv}
\end{equation}
 $S$ is the nuclear Schiff moment, $d_p$ is the proton EDM,
\begin{equation}
   X=\frac{2\pi}{3} \left[
     \frac{\partial}{\partial z}\rho_{\psi}(\vec{r})
      \right] _{x,y,z=0}\ ,
  \label{X}
\end{equation}
 $\rho_{\psi}(\vec{r})$ is the electronic density calculated from the
 electronic wave function $\psi$;
\begin{equation}
   d^M = 2 \sqrt{2}(d_p+d_N)
   \left(
   \frac{\mu}{Z}+ \frac{1}{2mc}
   \right)M\ ,
\vspace{3mm}
 \label{dm}
\end{equation}
 where
 $d_N$ is the nuclear EDM arising due to P,T-odd nuclear forces;
 $\mu$, $m$ and $Z$ are the magnetic moment, mass and charge of the Tl
 nucleus; $c$ is the velocity of light,
%
\begin{equation}
   M = \frac{1}{\sqrt{2}}\langle\psi |\sum_i
   \left(\frac{\vec{\alpha}_i \times \vec{\bf l}_i}{r_i^3}\right)_{z}
   |\psi\rangle\ ,
 \label{M}
\end{equation}
 $\vec{\bf l_i}$ is the orbital momentum for electron $i$;
 $\vec{\alpha}_i$ are its Dirac matrices.
%
 $R$ and $Q$ are
 factors determined by the {nuclear} structure of $^{205}$Tl:
\begin{eqnarray}
  R=\langle\psi_N({\oprn})|\sum_n(q_n/Z-\delta_{n,3s})r_n^2|
             \psi_N({\oprn})\rangle\ ,
 \label{R}\\
  Q=[3/5\langle\psi_N({\oprn})|\sum_n(q_n{\oprn})|
                 \psi_N({\oprn})\rangle -
 \nonumber
 \\
  ~~~1/Z\langle\psi_N({\oprn})|\sum_n(q_nr_n^2)|\psi_N({\oprn})\rangle
\\ \times
 \langle\psi_N({\oprn})|\sum_n(q_n{\oprn})/r_n^2|\psi_N({\oprn})
 \rangle]_{\vec{I}}\ ,
\nonumber
\label{Q}
\end{eqnarray}
 where $\psi_N({\oprn})$ is the nuclear wave function.

 Accounting for $H_{\rm eff}$ leads to different hyperfine splitting of TlF in
 parallel and antiparallel electric and magnetic fields. The level shift
  $h\nu = 4(d^V+d^M)\langle \vec{I}/I \cdot \vec{\lambda} \rangle$
 is measured experimentally (for the latest results on TlF see \cite{Cho:91}).

 The parameters $X$ of \Eref{X} and $M$ of \Eref{M} are determined by the
 electronic structure of the molecule. They were calculated recently for the
 ground $0^+$ state of TlF by Parpia \cite{Parpia:97} and by Laerdahl, Saue,
 {Faegri~Jr.}, and Quiney \cite{Laerdahl:97}, using the Dirac--Fock method with
 gaussian basis sets of large sizes (see Table \ref{result}).  (Below we refer
 to paper \cite{Quiney:98b} with the calculations presented in details and not
 to the brief communication \cite{Laerdahl:97} of the same authors.) In paper
 \cite{Petrov:02} the first calculation of $^{205}$TlF that accounts for
 correlation effects was performed (see also \cite{Dzuba:02} where the limit on
 the Schiff moment of $^{205}$Tl was recalculated).

\paragraph*{Details of calculations.}
 \label{sTlFcalc}

 In paper \cite{Petrov:02} 21-electron GRECP \cite{Mosyagin:97} for Tl was
 used.  The correlation spin-orbital basis set consisted of 26$s$, 25$p$,
 18$d$, 12$f$, and 10$g$ gaussian-type orbitals on Tl, contracted to
 $6s6p4d2f1g$. The basis was optimized in a series of atomic two-component
 GRECP calculations, with correlation included by the relativistic coupled
 cluster method \cite{Kaldor:99} with single and double cluster amplitudes; the
 average energy of the two lowest states of the atom was minimized. The basis
 set generation procedure is described in Refs.\ \cite{Isaev:00,Mosyagin:00}.
 The basis set was designed to describe correlation in the outer core $5s$ and
 $5p$ shells of Tl, in addition to the $5d$ and valence shells. While
 correlating the $5s$ and $5p$ shells may not be important for a majority of
 chemical and physical properties of the atom, it is essential for describing
 properties coming from the inner Tl region, including P,T-odd effects.  The
 ($14s9p4d3f$)/\-[$4s3p2d1f$] basis set from the ANO-L library \cite{MOLCAS} is
 used for fluorine.

 First, one-component SCF calculation of the $(1\sigma \dots
 7\sigma)^{14}\-(1\pi 2\pi3\pi)^{12}(1\delta )^4$ ground state of TlF is
 performed, using the spin-averaged GRECP for Tl which simulates the
 interactions of the valence and outer core ($5s5p5d$) electrons with the inner
 core [Kr]$4d_{3/2}^44d_{5/2}^64f_{5/2}^64f_{7/2}^8$.  This is followed by
 two-component RCC calculations, with only single (RCC-S) or with single and
 double (RCC-SD) cluster amplitudes. The RCC-S calculations with the
 spin-dependent GRECP operator take into account effects of spin-orbit
 interaction at the level of the one-configurational SCF-type method. The
 RCC-SD calculations include, in addition, the most important electron
 correlation effects.

 At the restoration stage, the nucleus was modeled as a uniform charge
 distribution within a sphere with radius
  $r_{\rm nucl} = 7.1\,{\rm fm} \equiv 1.34\times10^{-4}$ a.u.,
 whereas previous calculations \cite{Parpia:97,Quiney:98b} employed a spherical
 gaussian nuclear charge distribution (the root mean square radius in all
 calculations is 5.5 fm, in accord with the parametrization of Johnson and Soff
 \cite{Johnson:85}, and agrees with the experimental value 5.483 fm for the
 $^{205}$Tl nucleus \cite{Fricke:95}). The two equivalent [$15s12p12d8f$]
 numerical basis sets were generated for the restoration.

 The quality of the approximation for the two-center molecular spinors and,
 consequently, of the calculated properties increases with the value of $L_{\rm
 max}$. A series of calculations of the $M$ parameter was performed using
 \Eref{restoration} with basis functions including up to $p$, $d$ and $f$
 harmonics.  It is found (see Table \ref{result}) that accounting for only $s$
 and $p$ functions in the expansion determines $M$ with 90\% accuracy.  Because
 the contribution of $f$ functions is only about 0.3\% and amplitudes of higher
 harmonics on the nucleus are suppressed by the leading term $\sim
 r^{(j{-}1/2)}$, the error due to the neglect of spherical harmonics beyond $f$
 is estimated to be below 0.1\%.  Calculation of the $X$ parameter requires $s$
 and $p$ harmonics (see Ref.\ \cite{Quiney:98b}), although, strictly speaking,
 $d$ harmonics also give nonzero contributions.

 In the RCC calculations, the $X$ and $M$ parameters were calculated by the
 finite field method (e.g., see Refs.\ \cite{Kunik:71,Monkhorst:77}). The
 operator corresponding to a desired property (\Erefs{X} and
 \eref{M}) is multiplied by a small parameter $\lambda$ and added to the
 Hamiltonian. The derivative of the energy with respect to $\lambda$ gives the
 computed property. This is strictly correct only at the limit of vanishing
 $\lambda$, but it is usually possible to find a range of $\lambda$ values
 where the energy is almost linear with $\lambda$ and energy changes are large
 enough to allow sufficient precision. The quadratic dependence on $\lambda$ is
 eliminated in the present calculations by averaging absolute energy changes
 obtained with $\lambda$ of opposite signs.

\paragraph*{Results.}
 \label{sTlFres}

 Calculations were carried out at two internuclear separations, the equilibrium
 $R_e=2.0844$ \AA\, as in Ref.\ \cite{Parpia:97}, and 2.1 \AA, for comparison
 with Ref.\ \cite{Quiney:98b}. The results are collected in Table \ref{result}.
 The first point to notice is the difference between spin-averaged SCF values
 and RCC-S values, the latter include spin-orbit interaction effects. These
 effects increase $X$ by 9\% and decrease $M$ by 21\%.  The RCC-S function can
 be written as a single determinant, and results may therefore be compared with
 DHF values, even though the RCC-S function is not variational.
 GRECP/RCC-S/NOCR
 values of the $M$ parameter are indeed within 3\% and 1\% of difference with
 the corresponding DHF values \cite{Parpia:97,Quiney:98b} (Table \ref{result}).
 This agreement confirms the validity of the used approximations. In
 particular, freezing the inner core shells is justified, as inner core
 relaxation effects have little influence on the properties calculated here, a
 conclusion already drawn by Quiney et al.\ \cite{Quiney:98b}.

 Much larger differences occur for the $X$ parameter.  There are also large
 differences between the two DHF calculations for $X$, which cannot be
 explained by the small change in internuclear separation. The value of $X$ may
 be expected to be less stable than $M$, because it is determined by the
 derivative of the electronic density at the Tl nucleus and involves large
 cancellations \cite{Quiney:98b} between contributions of large and small
 components, each of them is about 20 times larger than their sum. Thus, a
 strong dependence of $X$ on the basis used may be expected. The DHF values
 collected in Table \ref{result} indeed show such dependence. Results obtained
 in Refs.\ \cite{Parpia:97} and \cite{Quiney:98b} with comparable even-tempered
 basis sets, $(28s28p12d8f)$ and $(28s28p14d8f)$, are rather close, differing
 by 340 a.u. Improving the Tl basis to $(34s34p16d9f)$ \cite{Quiney:98b}
 increases $X$ by 650 a.u.\ or 8\%.  Further improvement of the basis may be
 expected to yield even higher $X$ values.  The numerical basis functions
 obtained in atomic DHF calculations and used for the restoration are highly
 accurate near the nucleus, so that the GRECP/RCC-S/NOCR value for $X$, which
 is higher than that of Quiney\etal\ \cite{Quiney:98b}, seems reasonable.  The
 different nuclear models used in the present and DHF
 \cite{Parpia:97,Quiney:98b} calculations may also contribute to the
 disagreement in $X$, which is determined by the derivative of the electronic
 charge density at a single point, the Tl origin.  $M$ is affected by $\psi$ in
 a broader region, and is therefore far less sensitive to the nuclear model.

 The electron correlation effects are calculated by the RCC-SD method at the
 molecular equilibrium internuclear distance $R_e$. A major correlation
 contribution is observed, decreasing $M$ by 17\% and $X$ by 22\%.



 The hyperfine structure constants of Tl $6p_{1/2}^1$ and Tl$^{2+}$ $6s^1$,
 which (like $X$ and $M$) depend on operators concentrated near the Tl
 nucleus, were also calculated. The errors in the DF values are 10--15\%;
 the RCC-SD results are within 1--4\% of experiment. The improvement in $X$ and
 $M$ upon inclusion of correlation is expected to be similar.

 \squeezetable
\begin{table*}
\caption
 {Calculated $X$ \eref{X} and $M$ \eref{M} parameters (in a.u.)
 for the
 $^{205}$TlF ground state, compared with DHF values
 \protect\cite{Parpia:97,Quiney:98b}. GRECP/RCC-S results include spin-orbit
 interaction, and GRECP/RCC-SD values also account for electron correlation.
}
\medskip
\begin{ruledtabular}
\begin{tabular}{ll|ddd|r|dd|r}
                          & & \multicolumn{4}{c|}{$R_e=2.0844$ \AA} &
                          \multicolumn{3}{c}{$R=2.1$ \AA} \\
\hline
 \multicolumn{2}{c|}{Expansion}
 & $s,p$ & $s,p,d$ & $s,p,d,f$ & $s,p$ & $s,p$ & $s,p,d,f$ & $s,p$ \\
 \multicolumn{2}{l|}{}
 & \multicolumn{3}{c|}{$M$} & $X$ & \multicolumn{2}{c|}{$M$} & $X$    \\
 \multicolumn{2}{l|}{SCF(spin-averaged)}
             & 19.67 &  17.56 &  17.51 &  8967   & 19.52 &  17.43 &  8897   \\
\hline
 \multicolumn{2}{l|}{GRECP/RCC-S}
             & 16.12 &        &  13.84 &  9813   & 16.02 &  13.82 &  9726   \\
\hline
 DHF \cite{Parpia:97}&Tl:($28s28p12d8f$)  &15.61 & &     &  7743  &     &     &  \\
\hline
 DHF \cite{Quiney:98b}&Tl:($25s25p12d8f$) & & &   &  &&  13.64\footnotemark[1]& 8098 \\
      & Tl:($28s28p14d8f$)    &     &        &   &  &&  13.62\footnotemark[1]& 8089 \\
      & Tl:($31s31p15d8f$)    &     &        &   &  &&  13.66\footnotemark[1]& 8492 \\
      & Tl:($34s34p16d9f$)    &     &        &   &  &&  13.63\footnotemark[1]& 8747 \\
\hline
 \multicolumn{2}{l|}{GRECP/RCC-SD}
             &       &        &  11.50 &  7635   &       &        &

\end{tabular}
\end{ruledtabular}
\footnotetext[1]{ $M$
 is calculated in Ref.\ \cite{Quiney:98b} using two-center
 molecular spinors, corresponding to infinite $L_{max}$ in
 Eq.~(\ref{restoration}).
}

\label{result}
\end{table*}

\section{Calculations of~ $^{207}$PbO molecule.}
 \label{sPbO}

 It was noted previously that the experiments on the excited $a(1)$
 \cite{DeMille:00} or $B(1)$ \cite{Egorov:01} states of PbO having nonzero
 projection of total electronic momentum on the internuclear axis can be, in
 principle, sensitive enough to detect $d_e$ three or even four orders of
 magnitude lower than the current limit.  The knowledge of the effective
 electric field, $W_d$, seen by an unpaired electron is required for extracting
 $d_e$ from the measurements.  In papers \cite{Isaev:04,Petrov:04}, $W_d$ for
 the $a(1)$ and $B(1)$ states of the PbO molecule was calculated by the RCC-SD
 \cite{Kaldor:97,Landau:01c} and CI \cite{Buenker:74,\SODCIp} methods.  To
 provide an accuracy check in calculation of the electronic structure near the
 Pb nucleus the hyperfine constant, $A_{\parallel}$, was also calculated.

\paragraph*{Details of calculations.}
 \label{sPbOcalc}

 The 22-electron GRECP for Pb \cite{Mosyagin:97} is used at the first stage of
 the two-step calculations of PbO: the inner shells of the Pb atom ($1s$ to
 $4f$) are absorbed into the GRECP, and the $5s5p5d6s6p$ electrons and all the
 oxygen electrons are treated explicitly at the integral preparation part.  Two
 series of RCC-SD calculations with 10 (when freezing the $5s5p5d$ shells of
 lead employing the level shift technique \cite{Titov:99} and freezing the $1s$
 shell of oxigen after the RASSCF stage) and all 30 electrons treated
 explicitly were performed.  Only 10 electrons were treated explicitly in the
 CI calculations. For each series of calculations, correlation spin-orbital
 basis sets are optimized in atomic two-component GRECP/RCC calculations of Pb.
 The generated basis sets on Pb are ($14s18p16d8f$)/[$4s7p5d3f$] for
 30-electron and ($15s16p12d9f$)/[$5s7p4d2f$] for 10-electron calculations. The
 correlation-consistent ($10s5p2d1f$)/[$4s3p2d1f$] basis listed in the
 MOLCAS~4.1 library \cite{MOLCAS} is used for oxygen.  We found that the $f$
 basis function of oxygen has little effect on the core properties calculated
 here and it was not included into the RCC calculations to reduce expenses.

 The leading $\Lambda\Sigma$ coupling terms and configurations for the $a(1)$
 and $B(1)$ states are $^3\Sigma^+$ $\sigma_1^2\sigma_2^2\sigma_3^2 \pi_1^3
 \pi_2^1$ and $^3\Pi_1$ $\sigma_1^2\sigma_2^2\sigma_3^1 \pi_1^4 \pi_2^1$,
 correspondingly. The molecular orbitals used in the CI and RCC-SD calculations
 are obtained by the RASSCF method \cite{Olsen:88,MOLCAS} with the
 spin-averaged GRECP part \cite{Titov:99}, i.e.\ only scalar-relativistic
 interactions are taken into account in the RASSCF calculation.  The $a(1)$
 state is of practical interest first of all and, therefore, the lowest
 $^3\Sigma^+$ state was calculated at this stage.  Using the $C_{2v}$ point
 group classification scheme, two A$_1$ orbitals are in RAS1, six orbitals (two
 A$_1$, two B$_1$, and two B$_2$) in RAS2, and 50 (20 A$_1$, 6 A$_2$, 12 B$_1$,
 and 12 B$_2$) in RAS3 for 10-electron calculation.  2 A$_1$ orbitals are in
 RAS1, 6 orbitals (2 A$_1$, 2 B$_1$, and 2 B$_2$) in RAS2, and 41 (16 A$_1$, 5
 A$_2$, 10 B$_1$, and 10 B$_2$) in RAS3 for 30-electron calculation. No more
 than two holes in RAS1 and two particles in RAS3 are allowed.

 At the restoration stage the nucleus is modeled by a uniform charge
 distribution within a sphere of radius $r_{\rm nucl}=7.12\,{\rm
 fm}=1.35\times10^{-4}\,{\rm a.u}$. The root mean square radius of the nucleus
 is 5.52 fm, in accord with the parametrization of Johnson and Soff
 \cite{Johnson:85}, and agrees with the experimental value of 5.497 fm
 \cite{Fricke:95} for the $^{207}$Pb nucleus. Taking this value for the root
 mean square radius and a Fermi distribution for the nuclear charge changes
 $A_{\parallel}$ and $W_d$ by 0.1$\%$ or less. The equivalent basis sets
 generated are [$9s14p7d$] for 10 and [$6s7p5d$] for 30-electron calculation
 correspondingly.

 Then the RCC-SD method and the spin-orbit CI approach with the selected
 single- and double-excitations from some multiconfigurational reference states
 (``mains'') \cite{Buenker:74,Titov:01} are employed in the sets of different
 {$\Lambda$}S many-electron spin- and space-symmetry adapted basis functions

\paragraph*{Results.}
 \label{sPbOres}

 CI calculations \cite{Petrov:04} were performed at two internuclear distances,
 $R=3.8$ a.u., as is in RCC calculations, and $R=4.0$ a.u.\
 (in the RCC calculations \cite{Isaev:04} the only internuclear distance
 $R=3.8$ a.u.\ was used because of problem with convergence at larger
 distances).  The calculated values with the one-center expansion of the
 molecular spinors in the Pb core on either $s$, $s;p$ or $s;p;d$ partial waves
 are collected in \Tref{res1}.

 Let us consider the $5s,5p,5d$ orbitals of lead and $1s$ orbital of oxigen
 as the outercore ones and $\sigma_1$, $\sigma_2$, $\sigma_3$, $\pi_1$, $\pi_2$
 orbitals of the PbO (consisting mainly of $6s,6p$ orbitals of Pb and $2s,2p$
 orbitals of O) as valence.  Although in the CI calculations we take into
 account only the correlation between valence electrons, the accuracy attained
 in the CI calculation of $A_{\parallel}$ is much better than in the RCC-SD
 calculation.  The main problem with the RCC calculation was that the used
 Fock-space RCC-SD version was not optimal in accounting for the nondynamic
 correlations (see \cite{Isaev:00} for details of RCC-SD and CI calculations of
 the Pb atom), though the potential of the RCC approach for electronic
 structure calculations is very high in prospect, especially, in the framework
 of the intermediate Hamiltonian formulation \cite{Landau:01c}.

 Then we estimate the contribution from correlations of valence electrons with
 outercore ones (which also account for correlations between outercore pairs of
 electrons) as difference in the results of the corresponding 10- and
 30-electron GRECP/RCC calculations (see also \cite{Isaev:00} where this
 correction is applied to the Pb atom).  We designate such correlations in
 \Tref{res1} as ``outercore correlations''. When taking into account
 outercore contributions at the point $R=4.0$ a.u.\ we used the results of the
 RCC calculation at the point $R=3.8$ a.u.  Since these contributions are
 relatively small and because the correlations with the outercore electrons are
 more stable than correlations in the valence region when the internuclear
 distance is changed the used approximation seems us reasonable.
 At last, the linear extrapolation of the results of calculations to the
 experimental equilibrium distances, $R_e=4.06$ a.u.\ for $a(1)$
 \cite{Martin:88} and $R_e=3.91$ a.u.\ for $B(1)$ \cite{Huber:79} is applied.
 The final results are: $A_{\parallel} = -3826$~MHz, $W_d =
 -6.1{\cdot}10^{24}{\rm Hz}/(e\cdot {\rm cm})$ for $a(1)$ and $A_{\parallel} =
 4887$~MHz, $W_d = -8.0{\cdot}10^{24}{\rm Hz}/(e\cdot {\rm cm})$ for $B(1)$
 states.  The estimated error for the final $W_d$ parameter is 20\% for $B(1)$
 state.  For $a(1)$ the estimated error bounds put the actual $W_d$ value
 between 90\% and 130\% of the final value (for details see \cite{Petrov:04}).

\begin{table*}
\caption
{
 Calculated  parameters $A_{\parallel}$ (in MHz) and $W_d$ (in $10^{24}{\rm
 Hz}/(e\cdot {\rm cm})$) for the $a(1)$ and $B(1)$ states of $^{207}$PbO in the
 internuclear distances 3.8 and 4.0 a.u.  The experimental value of
 $A_{\parallel}$ in $a(1)$ is $-4113$\,MHz \cite{Hunter:02}. The preliminary
 experimental value of $A_{\parallel}$ for $B(1)$ is $5000 \pm 200$\,MHz
 \cite{Kawall:04a}.
}
\medskip
\begin{ruledtabular}
\begin{tabular}{lrrrrrrrrrrrrr}
State &\multicolumn{6}{c}{$a(1)$\ \ $\sigma_1^2\sigma_2^2\sigma_3^2 \pi_1^3
 \pi_2^1$ \ \  $^3\Sigma_1$}&
&\multicolumn{6}{c}{$B(1)$\ \ $\sigma_1^2\sigma_2^2\sigma_3^1 \pi_1^4
 \pi_2^1$\ \ $^3\Pi_1$}\\
 Parameters &\multicolumn{3}{c}{$A_{\parallel}$} &&\multicolumn{2}{c}{$W_d$}
& & \multicolumn{3}{c}{$A_{\parallel}$}&& \multicolumn{2}{c}{$W_d$}\\
 \cline{2-4}\cline{6-7}\cline{9-11}\cline{13-14}
 Expansion & s& s,p& s,p,d&& s,p& s,p,d& & s& s,p& s,p,d&& s,p&

s,p,d \\

\hline
\multicolumn{14}{c}{ }\vspace{-2mm} \\
\multicolumn{14}{c}{\bf Internuclear distance $R=3.8$ a.u. } \\
\multicolumn{14}{c}{ }\vspace{-2mm} \\

 10e-RASSCF

& -894 & -1505 & -1503 && 0.73 & 0.70 & & & & &&0.0 &0.0 \\

 30e-RASSCF

& -759 & -1705 & -1699 && 0.96 & 0.91 & & & &1900 &&0.0 &0.0 \\

\hline

 10e-RCC-SD \cite{Isaev:04}

&     &      & -2635  && -2.93& -3.05 & & & &3878&& -11.10 & -10.10 \\

 30e-RCC-SD \cite{Isaev:04}

&     &      & -2698  &&      & -4.10 & &    &     & 4081&& -9.10 &-9.70 \\

\it
outercore (30e-RCC-SD - 10e-RCC-SD)

&     &      & \it -63&&     & \it -1.05& &    &  & \it 203&&  &
\it 0.40  \\

\hline
10e-CI \cite{Petrov:04}
&     &       &     -3446 &&    &     -4.13 &&     &  &     4582&&       &
-10.64 \\
\bf FINAL \cite{Petrov:04}
&     &       &            &&       &           &&  & &         &&  &        \\
(10e-CI + outercore)
&     &       & \bf -3509 && & \bf -5.18  &&     & & \bf 4785 &&       & \bf -10.24 \\
\multicolumn{14}{c}{ }\vspace{-2mm} \\
\multicolumn{14}{c}{\bf Internuclear distance $R=4.0$ a.u. } \\
\multicolumn{14}{c}{ }\vspace{-2mm} \\

 10e-RASSCF

& -770 & -1384 & -1383 && 1.05   & 1.00   & & & & &&0.0 &0.0 \\

\hline

 10e-CI \cite{Petrov:04}
&     &       &     -3689  &&       &     -4.81 &&  & &    4762 &&  &    -7.18 \\
 {\bf FINAL} \cite{Petrov:04}
&     &       &            &&       &           &&  & &         &&  &        \\
 (10e-CI + outercore)\footnotemark[1]
&     &       & \bf -3752  &&       & \bf -5.86 &&  & &\bf 4965 &&  &\bf -6.78 \\

\end{tabular}
\end{ruledtabular}
\footnotetext[1]{
 It is assumed that the outercore contribution at the internuclear distance
 $R{=}4.0$ a.u.\ is approximately the same as is at $R{=}3.8$ a.u.
}
\label{res1}
\end{table*}


\section{Conclusions}
 \label{sConcl}

 The P,T-parity nonconservation parameters and hyperfine constants
 are calculated for the heavy-atom molecules which are of primary interest for
 modern experiments on search for the PNC effects.  It is found that accounting
 for the electron correlations is necessary for precise calculation of these
 properties.  The developed two-step (GRECP/NOCR) scheme of calculation of the
 properties heavily concentrated in atomic cores is a very efficient way to
 take account of these correlations with moderate efforts.  The results of the
 two-step
 calculations for the hyperfine constants differ less than 10\% from the
 corresponding experimental data.  The comparable level of accuracy is expected
 for the P,T-odd spin-rotational Hamiltonian parameters which can not be
 obtained experimentally.  The precision of the discussed
 calculations
 is limited by the current possibilities of the correlation methods and codes
 and not by the GRECP and NOCR approximations, despite the GRECP/NOCR method
 allows one to reduce seriously the expenses of the correlation treatment as
 compared to conventional Dirac-Coulomb(-Breit) approaches when using basis of
 molecular spin-orbitals instead of spinors
 etc.
%
 In turn, the attained accuracy is sufficient for a reliable interpretation of
 the results measured in the molecular PNC experiments.




\vspace{4mm}
 \paragraph*{Acknowledgments.}


 The present work is supported by the U.S.\ CRDF grant RP2--2339--GA--02
 and RFBR grant 03--03--32335.  A.P.\ is grateful to Ministry of
 education of Russian Federation (grant PD\,02--1.3--236) and to St.-Petersburg
 Committee of Science (grant PD\,03-1.3-60).  N.M.\ is also supported by the
 grants of Russian Science Support Foundation and the governor of Leningrad
 district.

\bibliography{}




\end{document}